 \documentclass[preprint2]{aastex}

\newcommand{\etal}{et al.}
\newcommand{\kms}{km~s$^{-1}$}

\shorttitle{SagDIG}
\shortauthors{Lee \& Kim}

\begin{document}


\title{Stellar Populations of 
       the Sagittarius Dwarf Irregular Galaxy} 


\author{Myung Gyoon Lee and Sang Chul Kim}
\affil{Department of Astronomy, Seoul National University, Seoul 151-742, 
Korea \\
email: mglee@astrog.snu.ac.kr,sckim@astro.snu.ac.kr }






\begin{abstract}

We present deep $BVRI$ CCD photometry of the stars in the dwarf irregular
galaxy SagDIG. 
The color-magnitude diagrams of the measured stars in SagDIG 
show a blue plume which consists mostly of young stellar
populations, and a well-defined red giant branch (RGB). 
The foreground reddening of SagDIG is estimated to be
$E(B-V)=0.06$.
The tip of the RGB is found to be at $I_{\rm TRGB} = 21.55 \pm 0.10$ mag. 
From this the distance to this galaxy
 is estimated to be $d = 1.18\pm 0.10 $ Mpc.
This result,  combined with its velocity information, shows that
it is a member of the Local Group.
The mean metallicity of the red giant branch is estimated to be
[Fe/H] $< -2.2$ dex. This shows that SagDIG is one of the most metal-poor
galaxies. 
Total magnitudes of SagDIG ($<r_H (= 107$ arcsec))
 are derived to be $B^T=13.99$ mag, $V^T=13.58$ mag, $R^T=13.19$ mag, and $I^T=12.88$ mag, 
and the corresponding absolute magnitudes
are $M_B=-11.62$ mag, $M_V=-11.97$ mag,  $M_R=-12.33$ mag,
and $M_I=-12.60$ mag.
Surface brightness profiles of the central part of SagDIG are approximately 
fit by a King model with a core concentration parameter
$c = \log (r_t / r_c ) \approx 0.6$, and those of the outer part follow
an exponential law with a scale length of 37 arcsec.
The central surface brightness is measured to be $\mu_B (0) = 24.21 $ 
mag arcsec$^{-2}$ and $\mu_V (0) =23.91  $ mag arcsec$^{-2}$.
The magnitudes and colors of the brightest blue and red stars in SagDIG (BSG and RSG) are measured to be, respectively, 
$<V(3)>_{BSG} = 19.89\pm 0.13$ mag, $<(B-V)(3)>_{BSG} = 0.08\pm 0.07$ mag, $<V(3)>_{RSG} = 20.39\pm 0.10$ mag, and $<(B-V)(3)>_{RSG} = 1.29\pm 0.12$ mag.
The corresponding absolute magnitudes are derived to be 
$<M_V(3)>_{BSG} = -5.66$ mag and $<M_V(3)>_{RSG} = -5.16$ mag, which
are about one magnitude fainter than those 
expected from conventional correlations with galaxy luminosity.

\centerline{[To appear in the Astronomical Journal in 2000]}
\centerline{[Also available from astro-ph/9910481]}

\end{abstract}




\section{Introduction}


SagDIG (UGCA 438, UKS 1927--177) is a faint dwarf irregular galaxy 
discovered by \citet{ces77} and \citet{lon78}.
In the discovery papers, \citet{ces77} and \citet{lon78} estimated, using the
brightest stars seen in the photographic plates, 
the distance to this galaxy to be 600 kpc and 1.1 Mpc, respectively. Since then
SagDIG has been considered to be a member of the Local Group.
It is known that SagDIG contains as much as $\sim 10^8 M_\odot$ of HI gas
\citep{ces77,lon78,you97}.
In addition, a few H II regions were detected by \citet{str91}.
The metallicity of an extended H II region in SagDIG has been estimated 
to be very low, 3\% of solar \citep{ski89}. 
To date there is limited information of stellar populations in SagDIG, which was
given briefly by \citet{coo88}. 
\citet{coo88} discovered about 25 carbon stars scattered beyond the body of the galaxy,
and estimated the distance to this galaxy using three methods: 
$(m-M)_0 \sim 25.3$ using the brightest blue stars,
$(m-M)_0 \sim 25.4$ using the $I$-band luminosity of the tip of the red giant branch,
and $(m-M)_0 \sim 25.0$ using the mean $I$-band magnitude of the carbon stars.

In this paper we present a study of stellar populations of SagDIG 
based on deep $BVRI$ CCD photometry.  
This paper is composed as follows.
Section 2 describes the observations and data reduction.
Section 3 investigates the morphological structure of SagDIG, and 
Section 4 presents the color-magnitude diagrams of SagDIG. 
Sections 5, 6 and 7 estimate the reddening, distance, and metallicity
of SagDIG, respectively.
Section 8 presents the surface photometry of SagDIG
and Section 9 discusses the group membership, 
stellar populations, and the brightest stars of SagDIG.
Finally, summary and conclusion are given in  \S10.

\section{OBSERVATIONS AND DATA REDUCTION}

$BVRI$ CCD images of SagDIG were obtained on the photometric night of 1994 
October 8 (UT) using the University of Hawaii 2.2 m telescope at Mauna Kea.
Table 1 lists the journal of the observations of SagDIG.
A grey scale map of the $V$-band CCD image of SagDIG 
is displayed in Fig. 1. The size of the field of view is $7'.5 \times 7'.5$
and the ($2\times 2$ binned) pixel scale of the CCD is 0.44 arcsec pixel$^{-1}$.

For the analysis of the data 
we have divided the field covered by our CCD images into three regions 
as shown in Fig. 1:
the C region which covers the central region ($101'' \times 70''$) of SagDIG, 
the I  region which covers the outer region ($238'' \times 119''$) 
of the galaxy, 
and the  F  region which represents
a control field with the same area of the C region plus I  region. 

Instrumental magnitudes of the stars in the CCD images were derived using
DoPHOT \citep{sch93}.
These magnitudes were transformed onto the standard system using the
standard stars observed during two nights 
including the same night \citep{lan92}. 
The transformation equations we derived from the photometry of the standard
stars are: 
$V = v - 0.075 (b-v) -0.118 X + constant$,
$(B-V) = 1.130 (b-v) -0.111 X + constant$,
$(V-R) = 0.970 (v-r) -0.030 X + constant$, and
$I = i +0.050 (v-i) - 0.082 X + constant$,
where upper cases and lower cases represent, respectively, the standard
system and instrumental system. $X$ represents the air mass.
The rms scatter of the solutions are 0.01 -- 0.02 mag.
The total number of stars which were measured at $V$ and at least one other
color in the CCD image is $\sim$8100. 
Table 2 lists $BVRI$ photometry of the measured
bright stars with $V<20.8$ mag in the C and I regions of SagDIG. 
The coordinates X and Y in Table 2 are given in units of 
pixel ( = 0.44 arcsec), and increase to the east and to the south, respectively.

\section{MORPHOLOGICAL STRUCTURE}

Fig.1 shows that there are many bright foreground stars in the CCD field, 
which is expected from the galactic position of SagDIG 
($l=21.06$ deg, and $b=-16.28$ deg). 
A strong concentration of much fainter stars than most foreground stars are
seen in the C region, which represent SagDIG.
The main body of the galaxy is mostly seen inside the C region, and  
is much elongated along  the east-west direction. It looks like a crescent
in the low contrast image.
However, the outer part of the galaxy seen
at the faint level extends out to the boundary of the I  region.
There are few bright stars which appear to be the members of SagDIG in the I  region.
 Instead there are many faint stars seen better in $R$ and $I$ images of the 
I  region, which are probably old red giant stars as shown later.
 A color map created by combining $BVR$ images shows that
most stars in the C region are impressively blue, 
while the faint stars in the I region are mostly yellow to red.
The mean ellipticity and position angle of the entire structure of SagDIG 
were roughly estimated to be $e=0.5$ and P.A. = 90 deg, respectively. 
Therefore SagDIG is an elongated galaxy, in the central region of which 
some bright stars are irregularly distributed. 
This kind of structure that some young stars are irregularly distributed
against the smooth background of old stellar populations is common among
irregular galaxies \citep{san71, lee93b, lee93, min96, lee99a, lee99b}.

\section{COLOR-MAGNITUDE DIAGRAMS}

In Figs. 2 and 3 we display $V$--$(B-V)$ and $I$--$(V-I)$ diagrams of 
 1500 measured stars in the C region, 1570 measured stars in the I  region, 
and 710 measured stars in the  F  region.
Note that the area of the (C+I)-region in the field is the same 
as that of the F  region so that we can estimate the contamination due to 
foreground stars by comparing directly the color-magnitude diagrams (CMDs)
of each region.

Several distinguishable features of the stars in SagDIG are seen in 
Figs. 2 and 3.
First, there is a remarkable blue plume of bright stars with $(B-V)<0.4$ and
$(V-I)<0.3$ 
in the C region (filled circles). 
There are much fewer blue stars in the I region (open circles)
 than in the C region. 
Comparison of the C and I regions and the F  region in Fig. 2 
shows that these bright blue stars  are mostly  members of SagDIG. There are
few stars with $(B-V)<0.4$ and $(V-I)<0.3$ in the F region.
The brightest end of the blue plume extends up to $V\approx 19.5$ mag at
$(B-V)\approx 0.1$. These stars are mostly massive stars which were formed
recently. 
Stars in the I  region are mostly fainter than $V \approx 21$ mag, 
much fainter than those bright stars in the C region.
This shows that there was much less star formation in the outer part
of SagDIG recently.

Secondly, Fig. 3 shows that there is a strong concentration of red stars fainter
than $I \approx 21.5$ mag in the C and I regions in addition to the blue plume
at $(V-I) \approx 0.0$ . 
Most of these red stars are probably old red giant branch (RGB) stars of SagDIG.
This feature is more clearly visible in the I region than in the C region.

\section{REDDENING}

We have estimated the foreground reddening toward SagDIG using the color-color
diagram of measured bright stars. Fig. 4 shows the $(B-V)$--$(V-I)$ diagram of
bright stars with $16<V<18$ mag (open squares) and $19<V<21$ mag (filled circles) and 
photometric errors smaller than 0.015 mag.
The stars with $19<V<21$ mag are fainter and farther than the stars
with $16<V<18$ mag, and they have photometric errors larger than the latter.
These bright stars are definitely foreground stars, as shown in the color-magnitude diagrams in the previous section.
Comparing the colors of these stars with the intrinsic relation of dwarfs
\citep{cou78} in Fig. 4, we have derived a reddening value of $E(B-V)=0.00\pm0.04$. This value is obviously a underestimate for the total
reddening toward the SagDIG, because these foreground stars are relatively
nearby. 
The mean magnitudes of the stars in the two groups shown in Fig. 4 
are $V \approx 17$ mag and $V \approx 20$ mag.
These magnitudes correspond roughly to the distances of 2.5 kpc and 10 kpc, respectively, if we assume these stars are dwarfs 
(for which $M_V \approx 5$ mag and $(B-V)_0 \approx 0.7$).

On the other hand, the reddening value for the position of SagDIG 
derived from the
extinction map of our Galaxy \citep{sch98} is $E(B-V)=0.125$. This value is
 much larger than that based on the color-color diagram.
If this value of $E(B-V)=0.125$ is adopted, then most measured stars will 
deviate significantly more from the intrinsic relation 
represented by the dashed line in Fig. 4.
Further studies of the reddening for SagDIG, for example, using $UBV$ photometry,
 are needed to resolve this  difference.
In this study we adopt the mean value of these two estimates: $E(B-V)=0.06$.
Extinction laws given by \citet{car89} are used to calculate the extinctions 
for other colors for the total-to-selective extinction 
ratio of $R_V = 3.2 $:
$A_V = 0.19$ mag and $A_I= 0.12$ mag.

\section{DISTANCE} 

We estimate the distance to SagDIG using the $I$ magnitude of the tip
of the RGB (TRGB), as described in Da Costa \& Armandroff (1990) and 
in Lee, Freedman, \& Madore (1993). 
The $I$ magnitude of the TRGB is estimated
using the $I-(V-I)$ diagram in Fig. 3 and 
the luminosity function of red giant stars.
Fig. 5 shows the $I$-band luminosity function of the measured red giant stars 
in the I  region, from which
the contribution due to field stars was subtracted. 

The I  region is more appropriate for measuring the magnitude of the TRGB
than the C region, because there are much less young bright stars 
in the I region
and because the crowding is much less severe in the I region than 
in the C region. 
In Fig. 5, as the magnitude increases, there is a sudden
increase at $I=21.55\pm0.10$ mag in the luminosity function of the
I  region, which corresponds to the TRGB seen in the color-magnitude 
diagram in Fig. 3. 
We have also used the edge detecting method described in Lee et al. (1993),
obtaining the same result as above.

Several stars brighter than the TRGB are mostly
asymptotic giant stars of intermediate age, also seen in other dwarf
galaxies \citep{lee99a, lee99b}. This is also consistent with the presence
of carbon stars of intermediate age discovered by \citet{coo88}. 

The mean observed color of the TRGB is estimated to be $(V-I)=1.35\pm 0.03$.
The bolometric magnitude of the TRGB is then calculated from
$M_{\rm bol}=-0.19{\rm [Fe/H]} - 3.81$.
Adopting a value for the metallicity of [Fe/H]  $\approx -2.4$  dex as 
estimated below, we obtain a value for 
the bolometric magnitude of $M_{\rm bol}=-3.35$ mag.
(The mean metallicity of SagDIG 
 is not certain as discussed in the next section.
Only an upper limit for the mean metallicity is determined: 
[Fe/H] $<-2.2$ dex.
However, the result for the distance estimate is not affected much by this
 because the TRGB method is insensitive to the metallicity.) 
The bolometric correction at $I$ for the TRGB is estimated 
to be BC$_I=0.57$ mag,
adopting a formula for the bolometric correction 
BC$_I$ = 0.881 -- 0.243$(V-I)_{\rm TRGB}$.
The intrinsic $I$ magnitude of the TRGB is then given by
$M_I= M_{\rm bol} - {\rm BC}_I = -3.93$ mag.
Finally the distance modulus of SagDIG is obtained:
$(m-M)_0 = 25.36 \pm 0.18$ mag 
(corresponding to a distance of $1.18\pm 0.10$ Mpc)
for an adopted extinction of $A_I=0.12$ mag.
This result is in excellent agreement with the result obtained using the
same method by \citet{coo88}, $(m-M)_0 = 25.4$.

\section{METALLICITY}

We have estimated roughly the mean metallicity of the RGB stars in SagDIG 
using  the $(V-I)$ color of the stars 0.5 mag fainter than the TRGB, $(V-I)_{-3.5}$.
This color is measured from the median value of the observed
colors of about 30 red giant
branch stars with $I=22.0\pm0.10$ mag, to be $(V-I)_{-3.5} = 1.17 \pm 0.02$ (m.e.) 
The reddening-corrected color is $(V-I)_{-3.5,0} = 1.10$ for the adopted reddening.
This value is $\approx 0.1-0.2$ mag bluer than the mean color of the RGBs in
other metal-poor dwarf galaxies such as UKS 2323--326 and DDO 210 \citep{lee99a,lee99b}.
This value is even bluer than the RGB of the metal-poor Galactic globular
cluster M15 with [Fe/H] = --2.17 dex, and is beyond the range 
of calibrator globular clusters 
for [Fe/H] -- $(V-I)_{-3.5}$ relation of the RGB.
This result shows that the mean metallicity of the RGB stars in SagDIG 
is lower than
[Fe/H] $=- 2.2$ dex.
The mean color $(V-I)_{-3.5,0} = 1.10 \pm 0.02$ leads to a value for the 
mean metallicity of [Fe/H] $= -2.8$ dex, 
if we use the [Fe/H] -- $(V-I)_{-3.5,0}$ relation as it is. 
However this estimate is based on the extrapolation of the calibration 
relation. 
In Fig. 6 we overlayed the locus of the red giant branch of M15 on the
$I-(V-I)$ diagram of the measured stars in SagDIG,
 shifted according to the distance and reddening of SagDIG.
If we adopt the zero reddening for SagDIG, the RGB of SagDIG will get
closer to that of M15, but still too blue to be in a good agreement
(in this case the metallicity of the RGB is derived to be [Fe/H] $= -2.4$ dex).
The broadening of the faint part of the RGB is mostly
due to the photometric errors, as shown by the error bars in Fig. 6.
The very low metallicity of the RGB obtained in this study is consistent
with the result derived from the spectroscopy of an extended H II region 
in SagDIG by Skillman et al. (1989). 
Skillman et al. (1989) found  
that the metallicity of the HII region is 3\% solar, which is
in the lowest among the HII regions in known dwarf galaxies.
Thus the mean metallicity of the RGB stars in SagDIG is very low,
and is in the lowest end in the metallicity of known 
dwarf galaxies 
\citep{lee93,lee95a,lee95b,mat98,lee99a,lee99b}.


\section{SURFACE PHOTOMETRY}

It is difficult to derive reliably the surface photometry of SagDIG,
because of the presence of
several bright foreground stars in the direction of SagDIG.
We have obtained  the surface photometry of SagDIG as follows.
First, we removed in the original CCD images the images 
of several 
bright stars which were obviously considered to be
foreground stars in the area of SagDIG, using IMEDIT in IRAF. 
Then we performed aperture photometry of 
SagDIG using the elliptical annular aperture with ellipticity of 0.5 and
P.A. = 90 deg derived in this study.
The value of the sky background was estimated from the mean intensity
of the field region.

The results of surface photometry of SagDIG are listed in Table 3 and 
are displayed in Fig. 7. In Table 3 $r_{\rm eff}$ presents 
the mean major radius of an annular aperture, 
and $r_{\rm out}$ represents the outer radius of an annular aperture.
Fig. 7 shows that the surface brightness profiles are almost
flat in the central region of the galaxy ($r < 40$ arcsec), 
and follow approximately an exponential law 
in the outer part ($r \ge 40$ arcsec).
We fit the $BVRI$ surface brightness profiles of the inner region of SagDIG 
with  single-mass isotropic King models \citep{kin66} and those 
of the outer region with an exponential law in Fig. 8.
Fig. 8 shows that the surface brightness profiles of the inner region
are roughly fit by a King model with a core concentration parameter
$c = \log ( r_t / r_c ) \approx 0.62 $, where $r_c$ and $r_t$ represent the core radius and tidal radius, respectively. The core radius of the inner region
is measured to be $r_c = 52\pm 3$ arcsec = $300\pm20$ pc.
Note that this value is much larger than that listed in Mateo (1998), 125 pc,
which is based on rough estimate from the old data by Longmore et al. (1978).
The core radius of SagDIG is found to be similar to those of other dwarf galaxies (Mateo 1998).
The $BVR$ surface brightness profiles of the outer region ($r>40$ arcsec)
are fit roughly by an exponential law 
with a scale length of $r_s=37 \pm 2$ arcsec = $210\pm10$ pc. 
Fig. 7(b) illustrates the surface color profiles of the galaxy. 
Fig. 7(b) shows that the mean color of the inner region of SagDIG, $(B-V) \approx 0.3$, is blue, which
is typical for dwarf irregular galaxies. 
Also the color profiles are bluer in the inner 1 arcmin region
than in the outer region, which is due to 
the presence of blue stars in the inner region of the galaxy.
The properties of the surface brightness and color profiles of SagDIG 
are similar to those of other dwarf galaxies
\citep{kor89,mat98,lee95a,lee95b,lee99a,lee99b}.

From the data of the surface photometry we have derived several
basic parameters of SagDIG as follows.
The standard radius and Holmberg radius of SagDIG are measured to be
$r_{25}$ = $58\pm2$ arcsec = 330 pc and $r_H$ = $107\pm3$  arcsec = 610 pc, respectively. 
The errors in these are measuring errors  for which the uncertainty in the
   photometry was not considered. 
Therefore these are underestimates of the true errors.
These radii correspond roughly to the boundaries of the C and I regions,
respectively, shown in Fig. 1.
Note that our value for the Holmberg radius is slightly larger than
the value Longmore \etal (1978) derived roughly from their photographic plate,
$r_H$ = 90 arcsec.

The central surface brightness of SagDIG is measured to be 
 $\mu_B (0) = 24.21$  mag arcsec$^{-2}$, 
$\mu_V (0) =23.91  $ mag arcsec$^{-2}$,
 $\mu_R (0) = 23.63 $  mag arcsec$^{-2}$, and
$\mu_I (0) =23.53  $ mag arcsec$^{-2}$.
The total magnitudes of SagDIG within $r_H$ are derived to be
$B^T=13.99$ mag, $V^T=13.58$ mag, $R^T=13.19$ mag, and $I^T=12.88$ mag, 
and the corresponding absolute magnitudes
are $M_B=-11.62$ mag, $M_V=-11.97$ mag,  $M_R=-12.33$ mag,
and $M_I=-12.60$ mag.
$B$ total magnitude of SagDIG derived in this study, $B^T=13.99$ mag, 
is much brighter than those given by Cesarsky \etal (1977) and
Longmore \etal (1978), $B^T=15.5$ mag and 14.7 mag, respectively.

\section{DISCUSSION}

\subsection{The Local Group Membership of SagDIG}

We have measured the distance to SagDIG to be $d=1.18\pm0.10$ Mpc
 from the
$I$-band magnitude of the TRGB, which is an accurate distance indicator
for resolved galaxies \citep{lee93,sal98}.
From this 
the distance of SagDIG from the barycenter of the Local Group is derived
to be 1.1 Mpc.
From the measured heliocentric velocity of this galaxy, $v=-77\pm5$ \kms
\citep{lon78,you97}, the velocity of SagDIG from the barycenter of
the Local Group is calculated to be $-12\pm 5$ \kms \citep{lee95a}.
This value is much smaller than the upper limit for the boundary
of the Local Group, 60 \kms. 
These results show that SagDIG is a member of the Local Group.

\subsection{Stellar Populations in SagDIG}

We have investigated roughly the properties of young stellar populations seen
in Fig. 2 using the theoretical isochrones.
In Fig. 9, we overlay, in the $V-(B-V)$ diagram of SagDIG, two isochrones
for the metallicities of $Z=0.001$ ([Fe/H] = -- 1.2 dex) and
$Z=0.0004$ ([Fe/H] = -- 1.6 dex) and ages of 
40 and 300 Myrs given by the Padova Group \citep{ber94}.
We choose first the isochrone with the lowest metallicity,  $Z=0.0004$, 
among the isochrones given by Bertelli et al. (1994), which are
represented by the dashed lines in Fig. 9.
However, the red parts of these isochrones do not match well with the observed
stars. The red stars in SagDIG appear to match better the isochones
with higher metallicty, $Z=0.001$ as shown by the solid lines in Fig. 9.
Fig. 9 shows 
(1) that the  brightest blue and red  stars in SagDIG match roughly
the blue and red loops of the isochrone with an age of about 40 Myrs, and
(2) that the  brightest main-sequence at $V\approx 21$  mag ($M_V \approx -4.5$ mag, about 1 mag below the brightest blue supergiants) indicates an age
of about 10 Myrs.
These results show that stars were formed as recently as 10 Myrs ago in 
SagDIG.
On the other hand, a well-developed RGB shown in the $I-(V-I)$ diagram
 of Fig. 3
shows that the bulk of the stars in SagDIG were formed 
before a few Gyrs ago, and a small number of AGB stars above the TRGB
were probably formed a few Gyrs ago.

Fig. 10 displays the $V$ luminosity function of the main-sequence stars
with $(B-V)<0.5$ in SagDIG. 
The luminosity function we derived is incomplete for the faint end, 
but is reasonably complete for $V<23$ mag. The bright part of the
luminosity function ($19<V<23$ mag, $-6.6 < M_V < -2.6$ mag) 
is approximately fit by a line 
with a logarithmic slope of $0.68\pm 0.12$. 
This value of the slope  is similar to those of
other irregular and spiral galaxies 
\citep{fre86,hoe86,lee99a,lee99b}.
 
The integrated H I flux of this galaxy was measured to be $32.6\pm1.2$ Jy \kms
\citep{you97}. 
With better data for the distance and luminosity obtained in this study, 
we derive an HI mass, $M_{HI} = 9.3 \times 10^6$ $M_\odot$, 
a $B$ luminosity, $L_B = 7.0 \times 10^6 $ $L_\odot$, 
and a ratio of $M_{HI} / L_B = 1.3~ M_\odot / L_\odot$ which is  
typical for dwarf irregular galaxies \citep{mat98}.

\subsection{The Brightest Blue and Red Stars in SagDIG}

Longmore \etal (1978) estimated roughly from the survey plate the magnitude
of the
brightest blue stars in SagDIG to be $B=19.6\pm 0.5$ mag, and used
this result to derive the distance to this galaxy, obtaining a value of
1.1 Mpc. With our photometry we can investigate in detail
the properties of the brightest stars in this galaxy.

In the $V-(B-V)$ diagram of the C and I regions of SagDIG (filled circles
and open circles, respectively)
 shown in Fig. 2,
it is obvious which stars are the three brightest blue stars in SagDIG,
because there are no foreground stars with $V<22$ mag and $(B-V)<0.4$
in the F region.
The three brightest blue stars (called BSG) in SagDIG 
are IDs 3124, 3039, and 3652, 
the photometry of which are listed in Table 2.
The mean observed magnitudes and colors of these three BSGs 
in SagDIG are derived to be, respectively, 
$<V(3)>_{BSG} = 19.61\pm0.21$ mag and $<(B-V)(3)>_{BSG} = 0.12\pm0.07$.
The corresponding absolute magnitude and color are
$<M_V(3)>_{BSG} = -5.94\pm0.21$ mag,  $<(B-V)(3)>_{BSG, 0} = 0.06\pm0.07$.
However, two of these stars are found to be compact objects showing $H_\alpha$
emission by Strobel et al. (1991): 
3124 (the first brightest) and 3652 (the third brightest) are 
listed as compact H II regions, SHK1 and SHK2, respectively, in Strobel et al. (1991). 
On the other hand, 
Skillman et al. (1989) detected H$\beta$ and H$\gamma$ emission
lines, but no [O III] line from the spectra of SHK1 and SHK2. They suggested
``that these objects are probably galactic (e.g. dwarf novae)''.
But the magnitude and color of these objects show strongly that these objects are the members of SagDIG. Further spectroscopy is needed
to identify the class of these objects.
If we omit 3124 and 3652 as the BSGs, then the three BSGs will be
3039, 3897 and 3694, which are adopted as the final BSGs in this study.
The mean observed magnitudes and colors of these three BSGs 
 in SagDIG are derived to be, respectively, 
$<V(3)>_{BSG} = 19.89\pm0.13$ mag and $<(B-V)(3)>_{BSG} = 0.08\pm0.07$.
The corresponding absolute magnitude and color are
$<M_V(3)>_{BSG} = -5.66\pm0.13$ mag,  $<(B-V)(3)>_{BSG, 0} = 0.02\pm0.07$.

There are three known H II regions in SagDIG, two of which are compact objects
as described above, and the remaining one is 
an extended H II region, SHK3 (Strobel et al. 1991). 
We searched for the exciting source of SHK3, finding a bright blue star
located in the center of SHK3: ID 6313. The observed magnitude and color of this
star are $V=21.50\pm0.01$, $(B-V)=-0.19\pm0.02$,
$(V-R)=-0.11\pm0.03$ and $(V-I)=-0.22\pm0.05$.
The absolute magnitude and color of this star are
$M_V = -4.05$ mag and  $(B-V)_0 =-0.25$, corresponding to O-type dwarf. 

On the other hand, it is not easy to select the three brightest red stars
(called RSG) in SagDIG, because the heavy contamination due to foreground
stars is severe in the red region of the CMDs. We just select the three
brightest red stars with $(B-V)>1.0$ in the C region. They are
IDs 4966, 3979 and 3640, which are listed in Table 2.
The mean observed magnitudes and colors of these three RSGs 
 are derived to be, respectively, 
$<V(3)>_{RSG} = 20.39\pm0.10$ mag and $<(B-V)(3)>_{RSG} = 1.29\pm0.12$.
The corresponding absolute magnitude and color are
$<M_V(3)>_{RSG} = -5.16\pm0.10$ mag,  $<(B-V)(3)>_{RSG, 0} = 1.23\pm0.12$.
It should be noted that these values are very uncertain for the brightest
red stars, because of uncertainty in the membership of these stars used.

The luminosity of the brightest stars in galaxies is known to be correlated
with the luminosity of the parent galaxies. 
Lyo \& Lee (1997) presented, from the analysis of 17 galaxies 
(with $M_B <-14$ mag) to which Cepheid distances are available,
calibrations for the relation between the magnitudes of the brightest stars and 
the magnitudes of the parent galaxies:
$<M_V(3)>_{BSG} = 0.30 M_B ({\rm gal}) - 3.02$ with $\sigma = 0.55$ mag,
and $<M_V(3)>_{RSG} = 0.21 M_B ({\rm gal}) - 3.84$ with $\sigma = 0.47$ mag.
Using these relations, 
we derive $<M_V(3)>_{BSG} = -6.51$ mag and $<M_V(3)>_{RSG} = -6.28$ mag for the
absolute magnitude of SagDIG as derived in the previous section. 
Thus the magnitudes of the brightest stars in SagDIG are 
about one mag fainter than those
expected from the relation for the bright galaxies.

\section{SUMMARY AND CONCLUSION}

We have presented a study of the stellar populations in the dwarf irregular
galaxy SagDIG
 based on deep $BVRI$ CCD photometry.
The primary results obtained in this study are summarized as follows and
the basic information of SagDIG is listed in Table 4.

(1) $BVRI$ color-magnitude diagrams of the stars in the $7'.5\times 7'.5$
 area of SagDIG have been presented.
 These color-magnitude diagrams  exhibit a blue plume, 
a well-defined RGB, and a small number of AGB stars with intermediate age.

(2) The tip of the RGB is found to be at $I=21.55\pm 0.10$  mag 
and $(V-I)=1.35\pm 0.03$ mag.
From this value we derive a distance modulus of SagDIG of $(m-M)_0=25.36\pm 0.18$ mag, 
and a distance of $1.18\pm 0.10$ Mpc. 
From this result and the systemic velocity of SagDIG, 
we conclude that SagDIG is a member of the Local Group.

(3) The mean observed color of the RGB at $M_I=-3.5$ mag is 
$(V-I)=1.17\pm 0.02$ mag.
 From this value we obtain roughly the mean metallicity of the RGB:
[Fe/H] $< -2.2 $ dex. The metallicity of the RGB in SagDIG
is the lowest among the metallicities of known dwarf irregular galaxies.

(4) The total magnitudes of SagDIG within $r_H$ (=107 arcsec) are derived to be
  $M_B = -11.62$ mag, $M_V = -11.97$ mag, $M_R = -12.33$ mag, 
and $M_I = -12.60$ mag.
The central surface brightness is measured to be 
$\mu_B (0) = 24.2 $ mag arcsec$^{-2}$ and 
$\mu_V (0) =23.9  $ mag arcsec$^{-2}$.
Surface brightness profiles of the central part of SagDIG are 
approximately 
fit by a King model with a core concentration parameter
$c = log (r_t / r_c ) \approx 0.62$, and those of the outer part
follow an exponential law with a scale length of $r_s = 37 arcsec = 210 $ pc. 

(5) The magnitudes of three brightest blue and red stars in SagDIG
are derived: 
$<M_V(3)>_{BSG} = -5.66 \pm0.13$ mag and $<M_V(3)>_{RSG} = -5.16\pm0.10$ mag,
 which are about one magnitude fainter than those 
expected from conventional correlations with galaxy luminosity.







\acknowledgments

The authors are grateful to Dr.~Yong-Ik Byun for his help in observations.
This research is supported by
the Ministry of Education, Basic Science Research Institute grant 
No.BSRI-98-5411 (to M.G.L.).




\appendix






\begin{deluxetable}{ccccc}
\tablecaption{JOURNAL OF OBSERVATIONS FOR SAGDIG \label{tbl-1}}
\tablewidth{0pt}
\tablehead{
\colhead{Filters} & \colhead{T$_{\rm exp}$}   & \colhead{Air mass}
& \colhead{FWHM} & \colhead{U.T.(Start)} } 
\startdata
$B$   &  900 s            & 1.35 & 1$''$.1 & 1994 Oct 8 05:54  \\
$V$   &  $2 \times 900$ s & 1.45 & 0$''$.9 & 1994 Oct 8 06:11  \\
$R$   &  600 s            & 1.61 & 0$''$.9 & 1994 Oct 8 07:03 \\
$I$   &  $3 \times 600$ s & 1.77 & 1$''$.0 & 1994 Oct 8 07:15 \\
\enddata
 
\end{deluxetable}

\begin{deluxetable}{ccccccc | ccccccc} 
\footnotesize 
\tablecaption{ PHOTOMETRY OF THE BRIGHT STARS WITH $V<20.8$ MAG IN  SAGDIG \label{tbl-2}}
\tablewidth{550pt}
\tablehead{
\colhead{ID} & \colhead{X(px)} & \colhead{Y(px)} & \colhead{$V$} & \colhead{($B$--$V$)} & \colhead{($V$--$R$)} & \colhead{($V$--$I$)} &
\colhead{ID} & \colhead{X(px)} & \colhead{Y(px)} & \colhead{$V$} & \colhead{($B$--$V$)} & \colhead{($V$--$R$)} & \colhead{($V$--$I$)} 
} 
\startdata
  2991&  435.7&  541.9&  19.75&   0.54&   0.40&   0.51&  4259&  386.5&  640.2&  20.60&   0.60&   0.42&   0.72\\ 
  3003&  654.2&  543.4&  20.62&   0.87&   0.55&   0.97&  4406&  365.8&  647.8&  20.61&   0.57&   0.36&   0.62\\ 
  3024&  530.4&  545.4&  20.10&   0.44&   0.31&   0.51&  4512&  340.8&  653.0&  19.58&   0.76&   0.50&   0.79\\ 
  3030&  511.1&  546.0&  19.25&   0.59&   0.40&   0.56&  4758&  548.9&  663.9&  20.16&  -0.03&  -0.02&  -0.22\\ 
  3039&  474.3&  546.9&  19.71&   0.02&   0.01&  -0.03&  4805&  528.1&  666.4&  20.38&  -0.02&  -0.10&   0.01\\ 
  3054&  417.5&  550.2&  20.48&   1.02&   0.71&   1.06&  4828&  499.1&  667.4&  20.74&   1.45&   0.85&   1.31\\ 
  3116&  394.8&  557.7&  19.56&   0.61&   0.38&   0.63&  4836&  286.2&  667.7&  20.53&   0.50&   0.32&   0.55\\ 
  3124&  482.0&  558.7&  19.31&   0.15&   0.26&   0.34&  4841&  319.3&  667.9&  20.37&   1.35&   0.85&   1.71\\ 
  3126&  743.6&  559.1&  20.38&   0.35&   0.22&   0.41&  4847&  516.6&  668.4&  20.75&   0.90&   0.62&   0.93\\ 
  3159&  691.1&  563.3&  19.03&   0.80&   0.55&   0.83&  4888&  512.7&  670.5&  20.75&   1.22&   0.69&   1.19\\ 
  3168&  423.7&  564.5&  19.79&   0.81&   0.54&   0.86&  4966&  559.9&  674.2&  20.27&   1.46&   0.97&   2.04\\ 
  3192&  632.8&  566.7&  19.02&   0.65&   0.43&   0.67&  5024&  526.8&  677.2&  20.44&  -0.03&   0.00&  -0.06\\ 
  3287&  474.1&  577.5&  20.78&   1.20&   0.84&   1.34&  5215&  592.2&  686.2&  19.32&   0.64&   0.41&   0.62\\ 
  3326&  454.2&  580.4&  19.96&   0.73&   0.46&   0.77&  5234&  664.9&  687.5&  20.34&   0.93&   0.64&   1.07\\ 
  3332&  312.6&  581.2&  19.98&   0.72&   0.51&   0.77&  5250&  402.6&  688.1&  20.10&   0.71&   0.46&   0.76\\ 
  3363&  354.9&  583.8&  19.45&   0.81&   0.47&   0.80&  5262&  493.1&  688.7&  20.73&   0.95&   0.64&   1.10\\ 
  3427&  313.5&  588.8&  19.79&   0.66&   0.42&   0.62&  5263&  567.5&  688.7&  20.00&   1.29&   0.90&   1.54\\ 
  3449&  687.7&  590.1&  20.12&   0.79&   0.49&   0.79&  5283&  282.8&  690.0&  19.64&   0.97&   0.68&   1.07\\ 
  3456&  296.7&  590.8&  20.32&   0.53&   0.37&   0.59&  5515&  282.1&  702.7&  19.10&   1.02&   0.68&   1.07\\ 
  3543&  369.8&  597.3&  19.96&   0.79&   0.49&   0.72&  5625&  241.1&  708.8&  20.39&   0.77&   0.54&   0.90\\ 
  3616&  696.2&  601.5&  19.34&   1.09&   0.77&   1.26&  5843&  463.2&  720.8&  19.86&   0.75&   0.51&   0.84\\ 
  3640&  369.6&  603.2&  20.52&   1.23&   0.86&   1.54&  5882&  369.7&  723.8&  19.32&   0.70&   0.46&   0.70\\ 
  3652&  512.2&  604.2&  19.79&   0.17&   0.79&   0.50&  5994&  645.3&  732.1&  20.30&   1.03&   0.75&   1.24\\ 
  3655&  401.9&  604.3&  19.18&   0.96&   0.72&   1.16&  6003&  411.8&  732.7&  19.03&   0.56&   0.38&   0.62\\ 
  3694&  483.0&  606.0&  20.00&   0.18&   0.13&   0.25&  6024&  359.1&  734.1&  19.66&   0.71&   0.45&   0.70\\ 
  3701&  682.2&  606.5&  20.66&   0.50&   0.35&   0.62&  6053&  221.5&  736.2&  19.62&   0.85&   0.56&   0.83\\ 
  3712&  660.6&  607.2&  20.40&   1.22&   0.84&   1.29&  6071&  722.1&  737.3&  20.59&   1.50&   0.97&   2.20\\ 
  3716&  248.9&  607.7&  19.59&   0.66&   0.41&   0.66&  6087&  355.1&  738.3&  19.58&   0.84&   0.54&   0.88\\ 
  3718&  388.0&  607.9&  19.31&   0.67&   0.45&   0.72&  6158&  320.8&  743.7&  19.85&   0.64&   0.42&   0.73\\ 
  3738&  667.3&  609.0&  20.75&   1.03&   0.75&   1.21&  6210&  224.5&  747.0&  20.59&   0.84&   0.59&   0.94\\ 
  3740&  566.1&  609.1&  20.28&   1.06&   0.77&   1.16&  6222&  260.2&  747.5&  19.29&   0.72&   0.46&   0.78\\ 
  3779&  484.1&  611.3&  20.33&   0.47&   0.37&   0.63&  6253&  731.8&  749.7&  20.56&   0.81&   0.55&   0.84\\ 
  3897&  519.8&  618.1&  19.96&   0.04&   0.03&  -0.11&  6294&  388.2&  752.5&  20.40&   0.79&   0.49&   0.72\\ 
  3906&  645.9&  618.8&  20.57&   0.81&   0.52&   0.86&  6371&  667.1&  757.3&  20.50&   0.77&   0.53&   0.88\\ 
  3966&  648.2&  621.9&  20.36&   0.94&   0.59&   1.05&  6377&  454.6&  757.8&  20.69&   1.33&   0.88&   2.06\\ 
  3974&  495.5&  622.4&  20.53&   0.02&  -0.01&  -0.13&  6421&  697.2&  762.0&  19.48&   0.64&   0.41&   0.73\\ 
  3979&  417.9&  622.5&  20.37&   1.20&   0.86&   1.52&  6540&  618.1&  771.9&  20.79&   1.15&   0.80&   1.32\\ 
  3990&  514.7&  623.2&  20.53&   0.04&   0.00&  -0.02&  6548&  555.1&  772.7&  20.24&   0.50&   0.44&   0.66\\ 
  4040&  302.7&  627.2&  20.13&   0.73&   0.44&   0.68&  6582&  654.9&  775.1&  20.36&   0.52&   0.38&   0.60\\ 
  4042&  518.2&  627.2&  20.22&   0.86&   0.58&   0.83&  6587&  738.5&  775.6&  20.57&   1.18&   0.85&   1.47\\ 
  4093&  420.0&  630.7&  19.12&   0.83&   0.54&   0.82&  6615&  339.0&  778.1&  19.76&   1.15&   0.80&   1.35\\ 
  4101&  485.1&  630.9&  20.62&  -0.03&  -0.01&  -0.09&  6621&  505.5&  778.8&  19.34&   0.53&   0.35&   0.59\\ 
  4112&  532.2&  631.9&  20.75&   0.01&  -0.03&  -0.05&  6704&  651.4&  788.2&  19.51&   0.46&   0.30&   0.53\\ 
  4121&  521.9&  632.1&  20.30&   0.03&   0.01&   0.00&  6757&  319.5&  793.8&  19.86&   0.72&   0.42&   0.81\\ 
  4130&  549.0&  632.5&  19.57&   0.76&   0.49&   0.78&  6811&  433.9&  799.4&  20.37&   1.40&   0.96&   1.79\\ 
  4168&  512.6&  634.9&  20.17&   0.05&   0.03&   0.04&  6831&  329.2&  803.1&  20.76&   0.94&   0.52&  -0.02\\ 
  4197&  712.6&  636.5&  19.56&   0.65&   0.37&   0.65&  6846&  251.8&  805.9&  20.58&   0.84&   0.55&   0.93\\ 
  4230&  567.7&  638.4&  20.60&   0.65&   0.42&   0.73&  6855&  644.3&  806.7&  19.82&   1.22&   0.81&   1.51\\ 
\enddata
 
\end{deluxetable}

\begin{deluxetable}{rcccc  rcccc}
\scriptsize
\tablenum{3}
\tablecaption{$BVRI$ SURFACE PHOTOMETRY OF SAGDIG.
 \label{tbl-3}}
\tablewidth{0pt}
 \tablehead{ 
\colhead{$r_{\rm eff}$ [$''$]} & \colhead{$\mu_B$} & \colhead{$\mu_V$} &  \colhead{$\mu_R$} &\colhead{$\mu_I$} &
\colhead{$r_{\rm out}$ [$''$]} & \colhead{$B$} & \colhead{$V$} & \colhead{$R$} & \colhead{$I$}  }
\startdata 
 6.2    & 24.21    & 23.91    & 23.61    & 23.54    &  8.8    & 18.79    & 18.49    & 18.22    & 18.10    \\
13.9    & 24.21    & 23.91    & 23.65    & 23.50    & 17.6    & 17.26    & 16.95    & 16.67    & 16.46    \\
22.4    & 24.25    & 24.00    & 23.74    & 23.55    & 26.4    & 16.37    & 16.06    & 15.75    & 15.48    \\
31.1    & 24.29    & 24.04    & 23.78    & 23.65    & 35.2    & 15.74    & 15.45    & 15.15    & 14.92    \\
39.8    & 24.37    & 24.13    & 23.88    & 23.77    & 44.0    & 15.24    & 14.95    & 14.65    & 14.44    \\
48.6    & 24.63    & 24.33    & 24.04    & 23.91    & 52.8    & 14.91    & 14.62    & 14.31    & 14.07    \\
57.4    & 24.96    & 24.66    & 24.39    & 24.07    & 61.6    & 14.65    & 14.34    & 14.01    & 13.76    \\
66.1    & 25.22    & 24.92    & 24.57    & 24.26    & 70.4    & 14.43    & 14.09    & 13.74    & 13.49    \\
74.9    & 25.44    & 25.14    & 24.78    & 24.37    & 79.2    & 14.29    & 13.93    & 13.57    & 13.28    \\
83.7    & 25.89    & 25.40    & 25.18    & 24.75    & 88.0    & 14.18    & 13.80    & 13.43    & 13.14    \\
92.5    & 26.08    & 25.77    & 25.39    & 25.12    & 96.8    & 14.09    & 13.70    & 13.32    & 13.01    \\
101.3   & 26.32    & 26.00    & 25.62    & 25.36    & 105.6   & 14.00    & 13.59    & 13.19    & 12.89    \\
110.1   & 26.64    & 26.31    & 25.90    & 25.67    & 114.4   & 14.00    & 13.52    & 13.12    & 12.80    \\
\enddata
\end{deluxetable}

\begin{deluxetable}{lcl}
\scriptsize
\tablenum{4}
\tablecaption{BASIC INFORMATION OF SAGDIG
 \label{tbl-4}}
\tablewidth{0pt}
 \tablehead{ \colhead{Parameter} & \colhead{Information} & \colhead{Reference} }
\startdata 
$\alpha_{2000}$, $\delta_{2000}$ & $19^h 29^m 59^s.0$ ,$-17^\circ 40' 41''$ &  1 \\
$l, b$ & 21.06 deg, --16.28 deg  & 1 \\
HI heliocentric radial velocity, $v_\odot$ & $-77 \pm 5$ km s$^{-1}$ & 1 \\
Foreground reddening &  $E(B-V) =0.06 \pm0.03$ mag & 2, 3 \\
Distance & $(m-M)_0 = 25.36\pm 0.18$, $d = 1.18\pm 0.10$ Mpc & 3 \\
Central surface brightness & $\mu_B(0) = 24.2$ , $\mu_V(0) = 23.9$ mag arcsec$^{-2}$ & 3 \\
Core radius & $r_c = 52\pm2$ arcsec = 300 pc & 3 \\
Standard radius  & $r_{25} = 58\pm2$ arcsec = 330 pc & 3 \\
Holmberg radius   & $r_H = 107\pm3$ arcsec = 610 pc & 3 \\
Scale length & $r_s = 37\pm2$ arcsec = 210  pc & 3 \\
Apparent total magnitude ($<r_H$) & $B=13.99$ mag, $V=13.58$ mag & 3 \\
Absolute total magnitude & $M_B = -11.62 $ mag, $M_V = -11.97$ mag & 3 \\
RGB metallicity, [Fe/H] & $< -2.2 $ dex & 3 \\
HI Flux & $32.6\pm1.2$ Jy km s$^{-1}$ & 4 \\
HI mass   & $M_{\rm HI} = 9.3 \times 10^6 M_\odot $  & 3, 4 \\
\enddata
\tablerefs{ (1) de Vaucouleurs et al. (1991) (RC3); (2) Schlegel, Finkbeiner \& Davis (1998); (3) This study; (4) Young \& Lo 1997 .}
\end{deluxetable}


\clearpage






\vfill
\begin{figure}[1]
\plotone{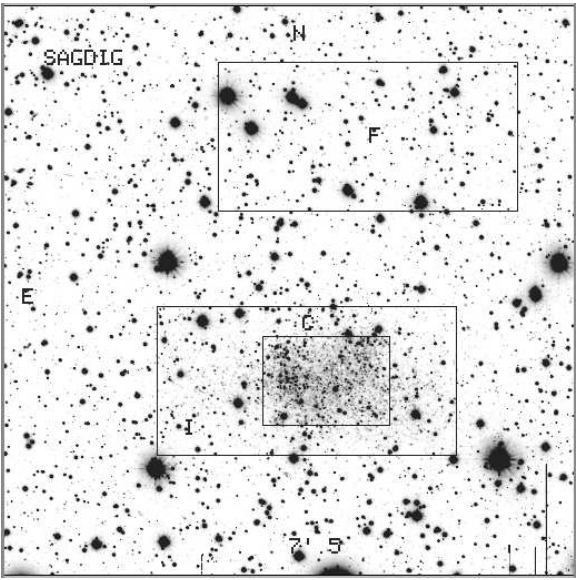}
\figcaption{
A greyscale map of the $V$-band CCD image of SagDIG. 
North is at the top and east is to the left.
The size of the field is $7'.5 \times 7'.5$. 
Regions labelled as C, I, and F represent,
 respectively,
the central region, the intermediate region and the control field region.
}
\end{figure}

\vfill
\begin{figure}[2] 
\plotone{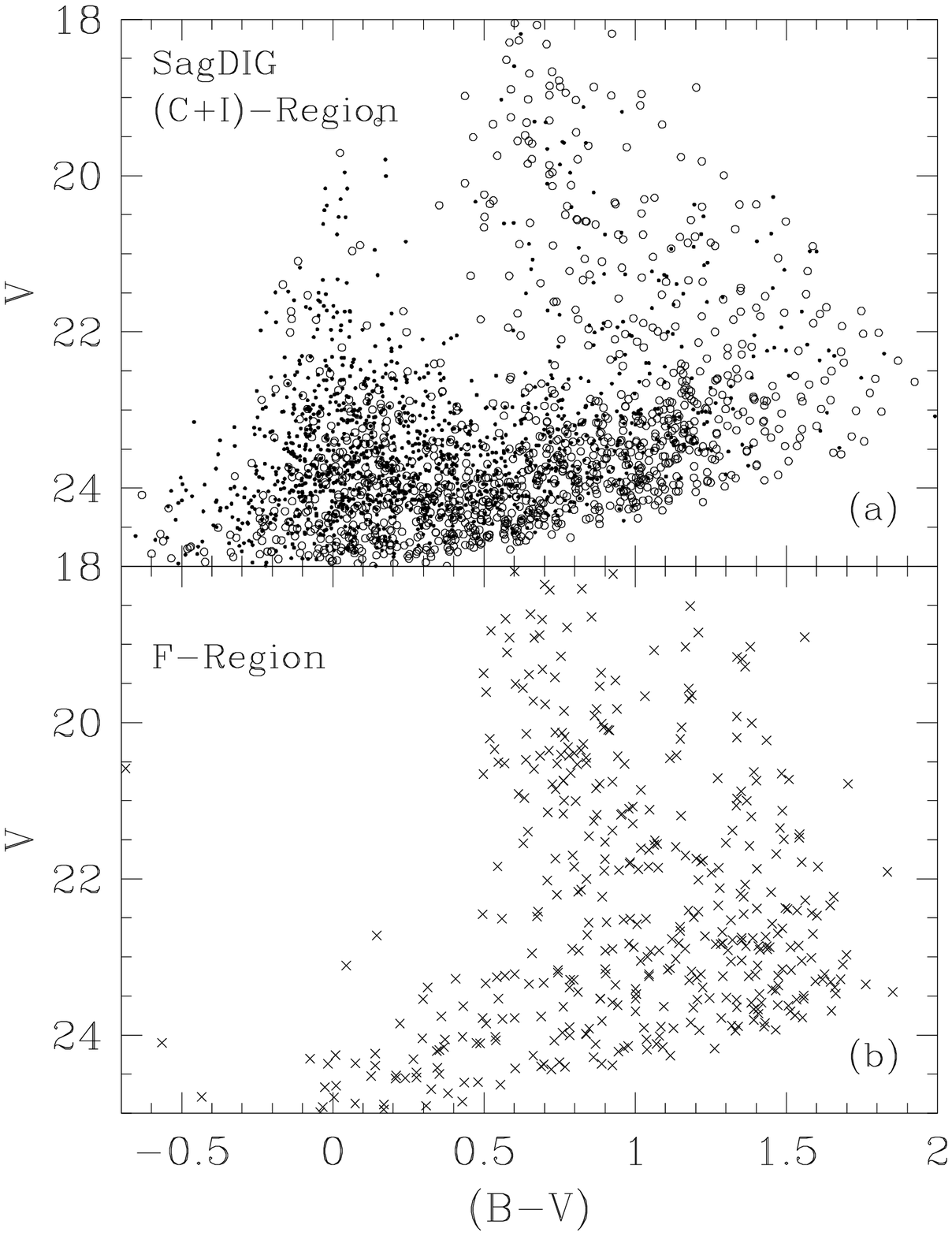}
\figcaption{(a) $V$--$(B-V)$ diagram of the measured stars in the C region plus I  region   of SagDIG. 
Filled circles and open circles represent the stars in the C region 
and I  region,  respectively.
(b) $V$--$(B-V)$ diagram of the measured stars in the F  region.}
\end{figure}
\begin{figure}[3] 
\plotone{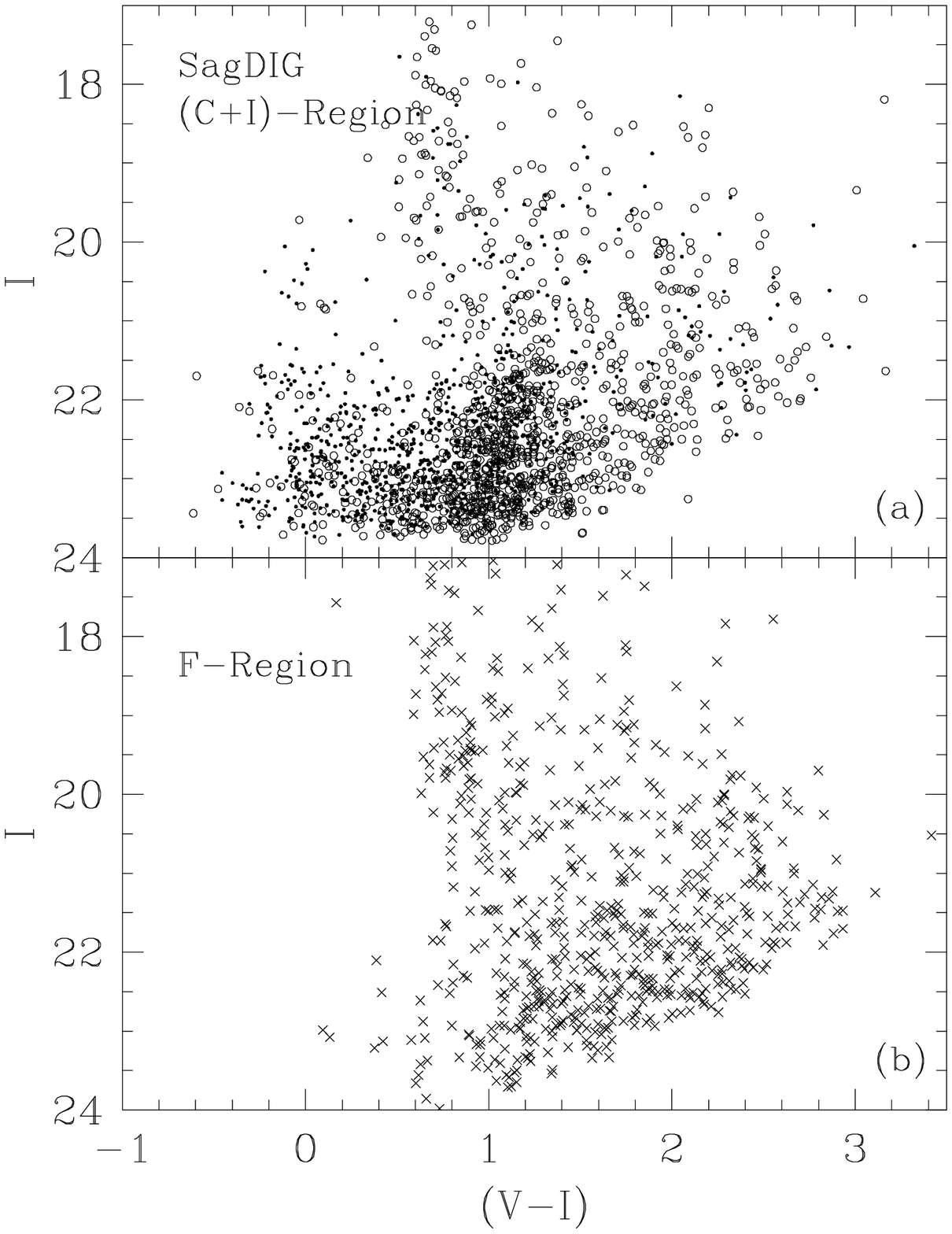}
\figcaption{
(a) $I$--$(V-I)$ diagram of the measured stars in the C region plus I  region
of SagDIG.
Filled circles and open circles represent the stars in the C region 
and I  region, respectively.
(b) $I$--$(V-I)$ diagram of the measured stars in the F  region.
}
\end{figure}
\begin{figure}[4] 
\plotone{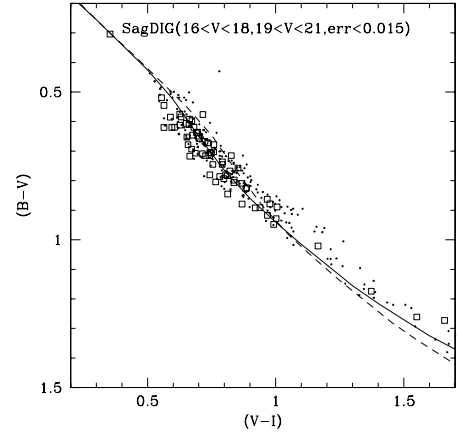}
\figcaption{
$(B-V)$--$(V-I)$ diagram for the bright stars with $16<V<18$ mag
(open squares), 
$19<V<21$ mag (filled circles) and photometric errors smaller than 0.015.
Solid and dashed lines represent the intrinsic relations for the dwarfs,
which are shifted according to the reddening of $E(B-V)=0.0$ and 0.125, respectively. 
Note that most stars are located along the line with $E(B-V)=0.0$.}
\end{figure}

\begin{figure}[5] 
\plotone{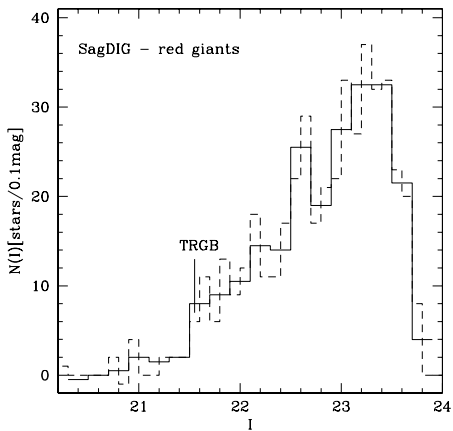}
\figcaption{
$I$-band luminosity function of the red giant branch stars in the I region.
The contribution due to field stars was subtracted from the luminosity
functions of the I  region. 
The solid line and dashed lines represent, respectively,
the luminosity functions derived using the bin size of 0.2 and 0.1 mag.}
\end{figure}

\begin{figure}[6] 
\plotone{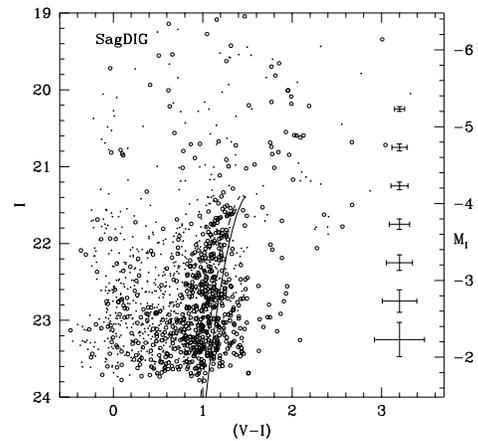}
\figcaption{
$I$--$(V-I)$ diagram of the measured stars in the I  region (open circles) and
C region (filled circles) of SagDIG in comparison with
the red giant branches of Galactic globular cluster M15. 
Solid and dashed curved lines show 
the locus of the giant branch of M15 ([Fe/H] $=-2.17$),
which is shifted to the reddening of $E(B-V)=0.06$ and 0.0, respectively,
and the distance of SagDIG.
The mean errors for the magnitudes and colors are illustrated 
by the error bars at the right.
Note that the mean red giant branch of SagDIG is even bluer 
than that of M15 with
$E(B-V)=0.0$.
}
\end{figure}

\begin{figure}[7] 
\plotone{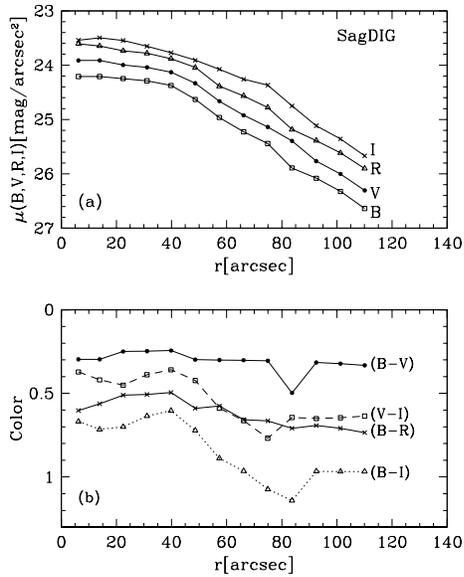}
\figcaption{ Surface photometry of SagDIG.
(a) Surface brightness profiles versus  radius along the major axis.
$B,V, R$ and $I$ magnitudes are represented by the open squares, 
filled circles, open triangles, 
and crosses. 
(b) Differential colors versus  radius along the major axis.
}
\end{figure}

\begin{figure}[8] 
\plotone{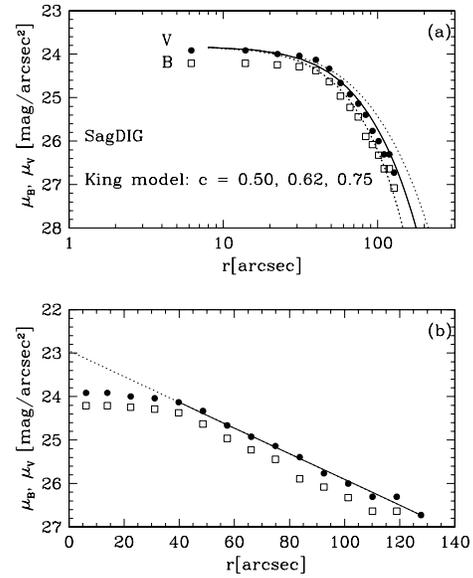}
\figcaption{Fits to the surface brightness profiles of SagDIG.
 Filled circles and open squares represent,
respectively, $V$ and $B$ surface brightness profiles.
(a) King model fitting.(b) Exponential law fitting.
}
\end{figure}

\begin{figure}[9] 
\plotone{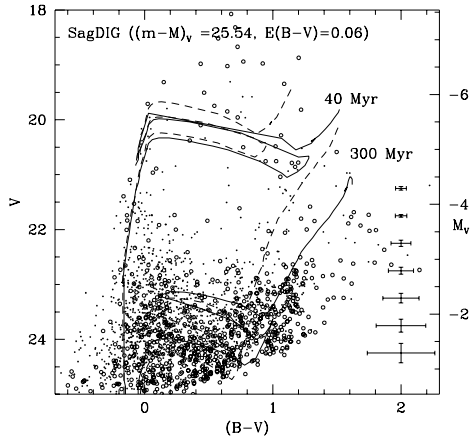}
\figcaption{
$V$--$(B-V)$ diagram of the measured stars in the C region (filled circles) and
I  region (open circles) of SagDIG in comparison with theoretical isochrones.
The solid lines represent the Padova isochrones for the metallicity of $Z=0.0004$
([Fe/H] = --1.6 dex),  $Z=0.001$
([Fe/H] = --1.2 dex) and ages of 40 and 300 Myrs.
The mean errors for the magnitudes and colors are illustrated by the error bars
at the right.
}
\end{figure}


\begin{figure}[10] 
\plotone{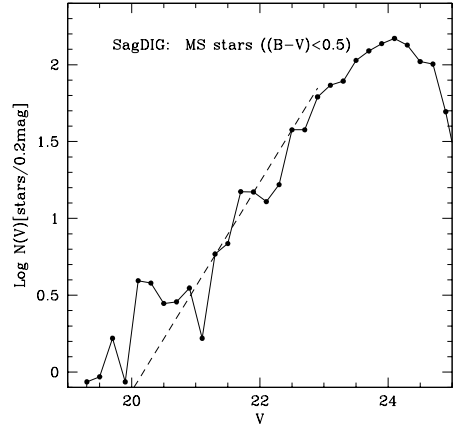}
\figcaption{ 
$V$-band luminosity function of the bright main-sequence stars with $(B-V)<0.5$
in SagDIG. The dashed line represents a linear fit to the data for the
range of $19<V<23.5$ mag.}
\end{figure}







\begin{thebibliography}{}


\bibitem[Bertelli \etal(1994)]{ber94} Bertelli, G., Bressan, A.,  Fagotto, F., Chiosi, C., \& Nasi, E. 
               1994, A\&AS, 106, 275 


\bibitem[Cardelli, Clayton, \& Mathis(1989)]{car89} Cardelli, J. A., Clayton, G. C., \& Mathis, J. S. 1989, ApJ, 345, 235

\bibitem[Cesarsky \etal(1977)]{ces77} Cesarsky, D., A., Laustsen, S., Lequeux, J.,
Schuster, H.-E., \& West, R. M. 1977, A\&A, 61, L31


\bibitem[Cook \& Aaronson(1988)]{coo88} Cook, K. H., \& Aaronson, M. 1988, in
The Extragalactic Distance Scale, eds. S. van den Bergh \& C. J. Pritchet, ASPCS, 4, 75


\bibitem[Cousins(1978)]{cou78} Cousins, A. W. J. 1978, MNASSA, 37, 62

\bibitem[Da Costa \& Armandroff(1990)]{dac90} Da Costa, G. S., \& Armandroff, T. E.
1990, AJ, 100, 162

\bibitem[ de Vaucouleurs \etal(1991)]{dev91} de Vaucouleurs, G., de Vaucouleurs, 
Paturel, G., and Fouque, P. 1991, Third Reference Catalog of Bright Galaxies
(Springer-Verlag, New York) (RC3)

\bibitem[Freedman(1986)]{fre86} Freedman, W. L. 1986, Luminous Stars and
Associations in Galaxies, edited by C. W. H. de Loore, A. J. Willis, \&
P. Laskarides, IAU Symposium No. 116 (Reidel, Dordrecht), p. 61

\bibitem[Hoessel(1986)]{hoe86} Hoessel, J. 1986, Luminous Stars and
Associations in Galaxies, edited by C. W. H. de Loore, A. J. Willis, \&
P. Laskarides, IAU Symposium No. 116 (Reidel, Dordrecht), p. 439


\bibitem[King(1966)]{kin66} King, I. 1966, AJ, 71, 276

\bibitem[Kormendy \& Djorgovski(1989)]{kor89} Kormendy, J., \& Djorgovski, S. 1989, ARA\&A, 27, 235

\bibitem[Landolt(1992)]{lan92} Landolt, A. U. 1992, AJ, 104, 340

\bibitem[Lee(1993)]{lee93b} Lee, M. G. 1993, ApJ, 408, 409

\bibitem[Lee \etal(1993)]{lee93} Lee, M. G., Freedman, W. L., \& Madore, B. F.
 1993, ApJ, 417, 553

\bibitem[Lee(1995a)]{lee95a} Lee, M. G. 1995a, Jour. Korean Astro. Soc., 28, 169  

\bibitem[Lee(1995b)]{lee95b} Lee, M. G. 1995b, AJ, 110, 1129  

\bibitem[Lee \& Byun(1999)]{lee99a} Lee, M. G., \& Byun, Y.-I. 1999, AJ, 118, 817 

\bibitem[Lee \etal(1999)]{lee99b} Lee, M. G. et al. 1999, AJ, 118, 853 

\bibitem[Longmore et al.(1978)]{lon78} Longmore, A. J., Hawarden, T. G., Webster,
 B. L., Goss, W. M., \& Mebold, U. 1978, MNRAS, 183, 97P


\bibitem[Lyo \& Lee(1997)]{lyo97} Lyo, A.-R., \& Lee, M. G. 1997, Jour. Korean 
Astro. Soc., 30, 27

\bibitem[Mateo(1998)]{mat98} Mateo, M. 1998, ARA\&A, 36, 435


\bibitem[Minniti \& Zijlstra(1996)]{min96} Minniti, D., \& Zijlsta, A. A. 1996,
ApJ, 467, L13 


\bibitem[Salaris \& Cassisi(1998)]{sal98} Salaris, M., \& Cassisi, S. 1998, MNRAS, 298, 166

\bibitem[Sandage(1971)]{san71} Sandage, A. 1971, in Nuclei of Galaxies, 
ed. D.J.K. O'Connel (Amsterdam, North-Holland), 601

\bibitem[Schechter, Mateo, \& Saha(1993)]{sch93} Schechter, P., Mateo, M.
, \& Saha, A. 1993, PASP, 105, 1342

\bibitem[Schlegel, Finkbeiner, \& Davis(1998)]{sch98} Schlegel, D. J., Finkbeiner, D. P., \&
     Davis, M. 1998, ApJ, 500, 525

\bibitem[Skillman, Terlevich, \& Melnick(1989)]{ski89} Skillman, E., D.,
Terlevich, R., \& Melnick, J. 1989, MNRAS, 240, 563



\bibitem[Strobel, Hodge, \& Kennicutt(1991)]{str91}Strobel, N. V., Hodge, P., \& Kennicutt, Jr., R. C. 1991, ApJ, 383, 148



\bibitem[Young \& Lo(1997)]{you97} Young, L. M., \& Lo, K. Y. 1997, ApJ, 490, 710

\end{thebibliography}
\end{document}